\documentclass[a4paper,11pt]{article}
\usepackage{graphicx} 
\usepackage{authblk}
\usepackage{hyperref}
\usepackage{siunitx}
\usepackage{xcolor}
\usepackage{tikz}

\usepackage{caption}
\usepackage{subcaption}
\usepackage{authblk}

\usepackage{lineno}

\usepackage[a4paper, left=3.5cm, right=3.5cm]{geometry}

\title{\bfseries The ATLAS Trigger System}

\author{
    {\large Leonardo Toffolin$^{1,2,3}$ on behalf of the ATLAS Collaboration \thanks{\copyright ~Copyright 2025 CERN for the benefit of the ATLAS Collaboration. CC-BY-4.0 license.}}\\
    {\normalsize
    $^{1}$ University of Trieste, Italy\\
    $^{2}$ INFN Trieste, Gruppo Collegato di Udine, Italy\\
    $^{3}$ CERN, Geneva, Switzerland}
}

\date{Proceedings of the 32$^{\rm nd}$ International Symposium on Lepton Photon Interactions at High Energies (LP2025),\\
25–29 August 2025, Madison, WI, USA}

\begin{document}
\maketitle

\begin{abstract}
    The ATLAS Trigger system is a key component of the ATLAS experiment at the CERN Large Hadron Collider (LHC), designed to reduce the event rate from the 40~MHz proton--proton bunch crossing frequency to an output suitable for offline storage and analysis. During Run-3 (2022--2026), major upgrades were implemented in both the hardware-based Level-1 (L1) Trigger and the software-based High Level Trigger (HLT), to cope with increased luminosity and pile-up conditions. This paper summarises the main features of the ATLAS Trigger system, its performance in Run~3, and its role in enabling precision measurements and new physics searches.
\end{abstract}

\section{Introduction}

The ATLAS Trigger system~\cite{ATLAS,atlas_trigger2024} provides a multi-level event selection which takes inputs from the various subdetectors of the ATLAS Experiment. It is designed to efficiently retain interesting physics events while reducing the total data rate by around a factor of $10.000$. At the LHC, proton–proton ($pp$) collisions occur every 25~ns, generating an immense data flow from the ATLAS subdetectors which cannot be fully recorded. To ensure that only the most relevant information is available for reconstruction and analysis, the event rate is reduced by selecting and storing events that satisfy predefined trigger conditions.

\begin{figure}
    \centering
    \includegraphics[width=1.0\linewidth]{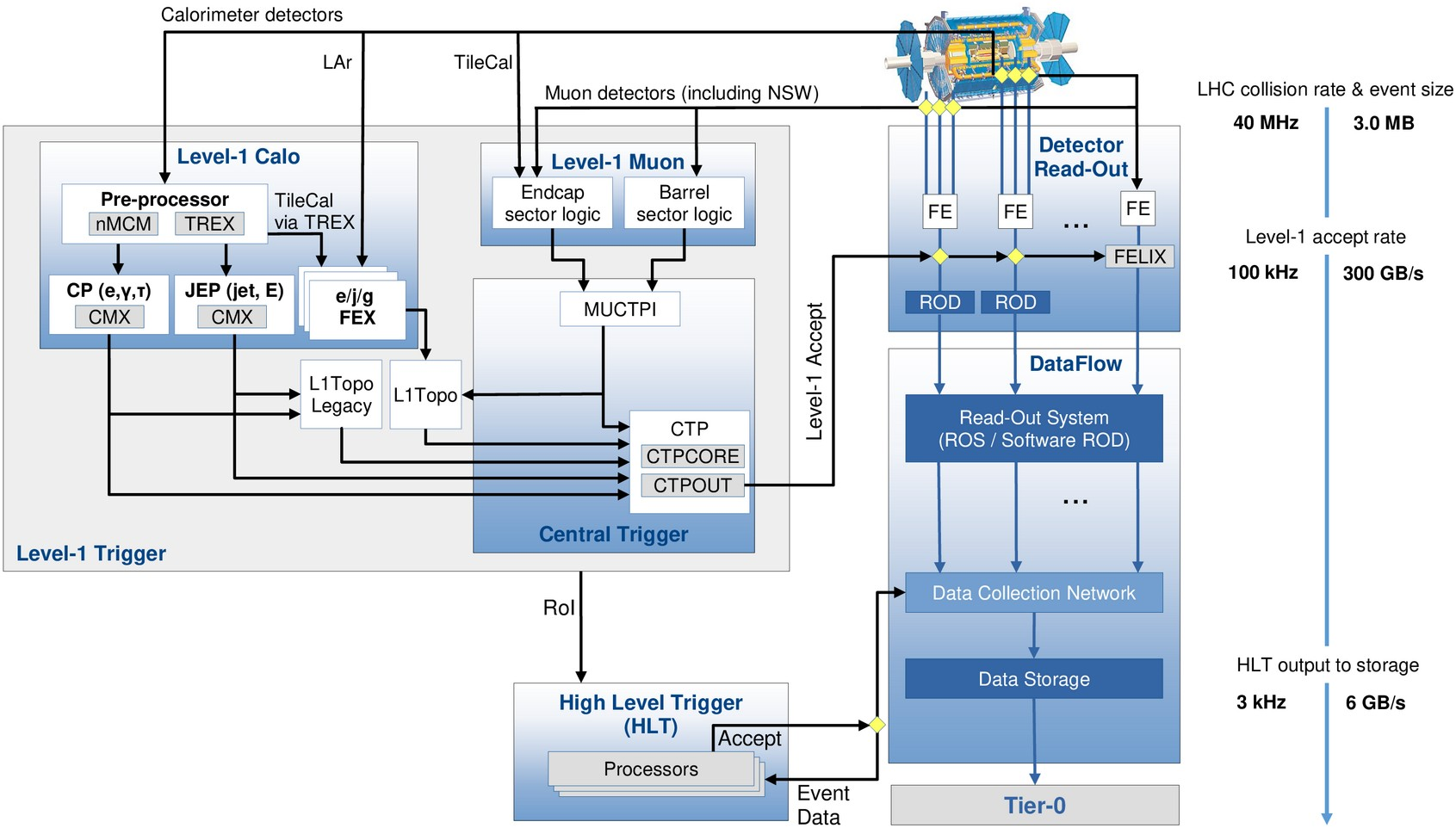}
    \caption{Architecture of the ATLAS Trigger system~\cite{atlas_trigger2024}.}
    \label{fig:Trigger architecture}
\end{figure}

The ATLAS Trigger system (Figure~\ref{fig:Trigger architecture}) operates in two main stages: the hardware-based Level-1 (L1) Trigger, and the software-based High Level Trigger (HLT). The L1 system uses custom electronics to identify physics objects with coarse granularity information from the calorimeters and muon detectors, reducing the rate to about $100$~kHz. The HLT then performs more refined event reconstruction using full detector granularity information, further reducing the rate of the main physics stream to about $3$~kHz, suitable for offline processing. At the beginning of Run~3 of the LHC, which started in 2022 with a centre-of-mass energy of $\sqrt{s} = 13.6~\text{TeV}$, the ATLAS Trigger system underwent significant upgrades to maintain high efficiency and flexibility under higher instantaneous luminosity conditions and a greater pile-up average value $\mu \sim 60$, to be compared with $\mu \sim 34$ for the 2018 $pp$ data taking.

\section{Level-1 Trigger System}

The L1 Trigger~\cite{l1calo_public,l1muon_public} consists of several subsystems: the L1 Calorimeter Trigger (L1Calo), the L1 Topological Trigger (L1Topo), the L1 Muon Trigger (L1Muon), and the Central Trigger Processor (CTP). The L1Muon sends information to the CTP through the Muon-to-CTP Interface (MuCTPI).


\paragraph{L1 Calorimeter Trigger}

\begin{figure}[ht]
    \centering
    \includegraphics[width=0.6\textwidth]{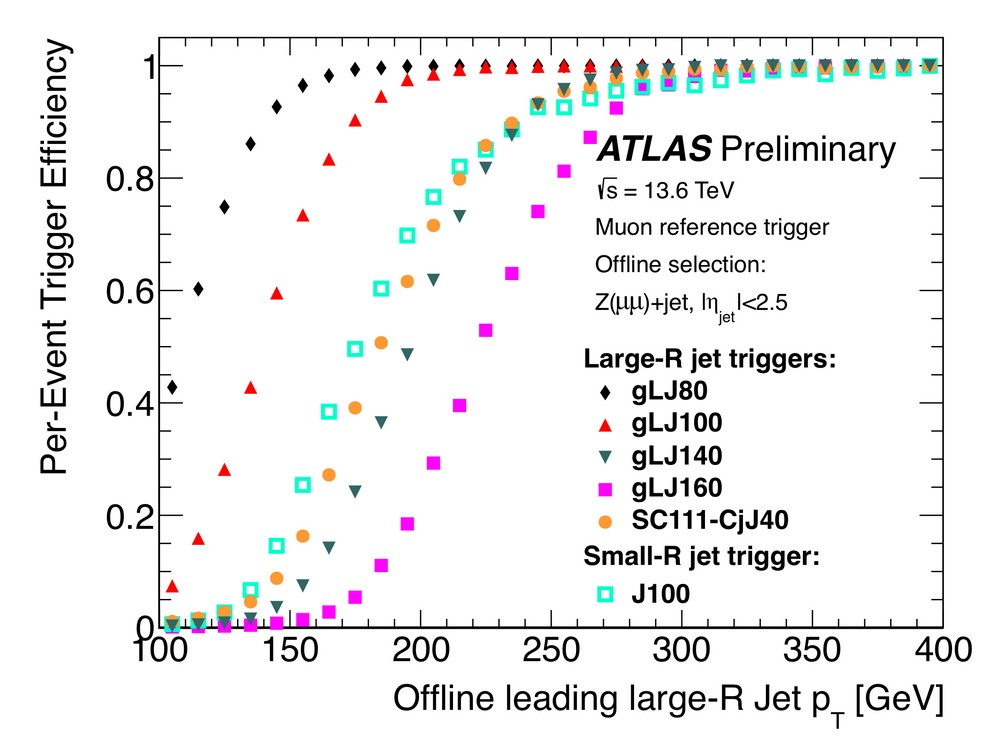}
    \caption{Efficiency of the L1 single large-$R$ jet trigger in Run-3. The large-$R$ jet variants present a sharper turn-on and better plateau efficiency than the legacy J100 (small-$R$ jet) item~\cite{l1calo_public}.}
\label{fig:L1Calo}
\end{figure}

The L1Calo system identifies electromagnetic (EM) objects, hadronic taus, and jets, and computes global quantities such as missing transverse energy ($E_{\mathrm{T}}^{\mathrm{miss}}$). The major improvements introduced for Run-3 include a finer granularity from the Liquid Argon (LAr) calorimeter readout, and new ATCA-based Feature Extractor (FEX) modules:
\begin{itemize}
    \item eFEX for electron, photon, and tau identification, featuring sophisticated clustering and isolation algorithms;
    \item jFEX for jet and $E_{\mathrm{T}}^{\mathrm{miss}}$ reconstruction;
    \item gFEX for global event quantities and complex topology, such as identification of large-radius jets.
\end{itemize}
The L1 efficiency of identification of large-radius jets in Run-3 is shown in Figure~\ref{fig:L1Calo}.

\paragraph{L1 Muon Trigger}

\begin{figure}[ht]
    \centering
    \begin{subfigure}[c]{0.508\textwidth}
    \includegraphics[width=\textwidth]{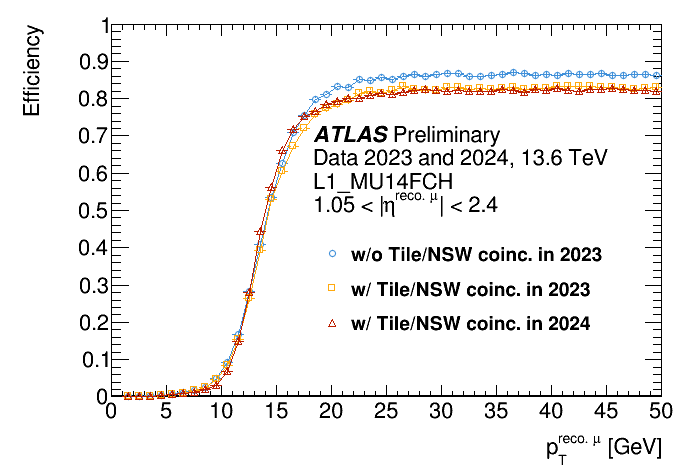}
    \caption{}
    \label{fig:L1Calo_subA}
    \end{subfigure}
    \hfill
    \centering
    \begin{subfigure}[c]{0.483\textwidth}
    \includegraphics[width=\textwidth]{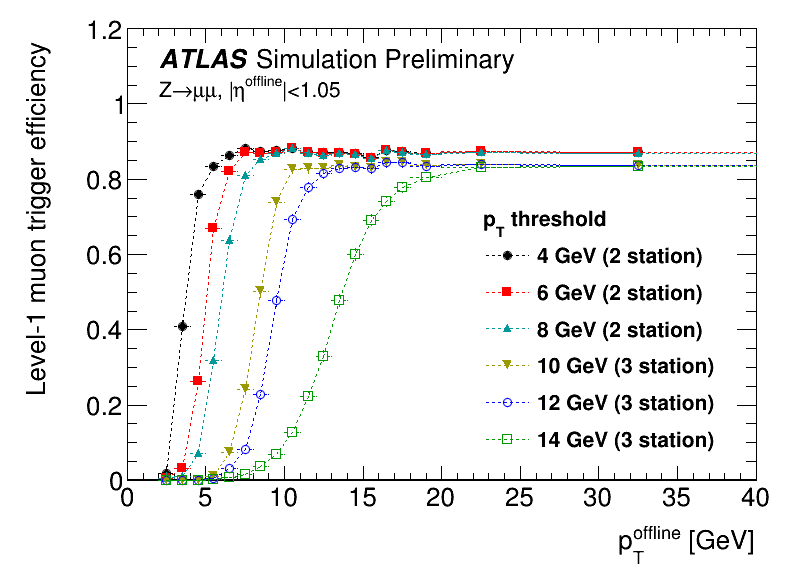}
    \caption{}
    \label{fig:L1Calo_subB}
    \end{subfigure}
    \caption{(a) Efficiency of the L1 muon trigger with 2023 and 2024 Run-3 data~\cite{l1muon_public}. (b) Efficiency of the L1 single muon trigger in Run-3~\cite{l1muon_public}.}
\label{fig:L1Muon}
\end{figure}

The L1Muon system combines signals from various muon trigger detectors: Resistive Plate Chambers (RPCs) in the barrel region, Thin Gap Chambers (TGCs) in the endcaps, and the New Small Wheel (NSW) detectors covering $1.3 < |\eta| < 2.7$. The NSW, composed of MicroMegas (MM) and small-strip TGC (sTGC) detectors, was introduced at the start of Run~3 to improve fake-muon rejection. The muon detectors are in coincidence with the Tile calorimeter. Overall, the inclusion of the NSW/TGC coincidence logic has reduced the L1 muon trigger rate by approximately 14 kHz, maintaining high efficiency across the full acceptance. The efficiency of the L1 muon trigger with Run-3 data is shown in Figure~\ref{fig:L1Muon}: in particular, Figure~\ref{fig:L1Calo_subA} illustrates the effect of the inclusion of the coincidence between the Tile calorimeter and the NSW.

\paragraph{L1Topo and CTP}

The L1 Topological Trigger (L1Topo) receives trigger objects from L1Calo and L1Muon to perform topological selections, such as angular separation ($\Delta R$) between reconstructed physics objects, enhancing background rejection for complex signatures. The Central Trigger Processor (CTP) receives inputs from all L1 systems, applies prescales and deadtime vetoes, and forms the final L1 Accept signal.

\section{High Level Trigger}

The HLT software~\cite{atlas_trigger2024} refines event selection using full detector granularity information and algorithms close to their offline version. During Run-3, the system underwent major software modernisation:
\begin{itemize}
    \item migration to a fully multithreaded framework;
    \item significant speed-up of track reconstruction, enabling full-scan operation;
    \item enhanced triggers for pile-up sensitive or unconventional topologies.
\end{itemize}
In the HLT, reconstruction algorithms run in a large farm of around $60.000$ processor cores, and a decision is formed typically within 300 ms. The HLT reduces the output event rate further down to about $3$~kHz for the physics main stream (Section~\ref{sec:Data streams}) at peak times. Once an event is accepted by the HLT, \textit{i.e.} once it passes the trigger selection, the Sub-Farm Output (SFO) sends the data to permanent storage for offline reconstruction and exports the data to the Tier-0 facility~\cite{Tier-0} at CERN’s computing centre. During the time that it takes the HLT to make a decision, the event information is retained in the Read-Out-system Buffer (ROB).


\section{Data streams and operations}
\label{sec:Data streams}

The Trigger system routes accepted events into multiple \emph{data streams}, which are collection of events or event fragments in the same dataset. Each stream serves a specific purpose:
\begin{itemize}
    \item \textbf{Main stream}: standard physics analyses and data quality.
    \item \textbf{Express stream}: prompt reconstruction for quick feedback.
    \item \textbf{Delayed streams} (\textit{e.g.}\ $B$-physics, vector boson fusion processes): events processed outside of data-taking periods (\textit{i.e.} LHC shutdowns and technical stops), when a more free computing capacity at the Tier-0 is available;
    \item \textbf{Calibration streams}: high-rate, reduced-content events for detector calibration.
    \item \textbf{TLA stream}: reduced event content for Trigger-Level Analysis. Such smaller event size (4.5~kB) with respect to the physics main stream (1.5~MB) allows more events to be saved, with a rate up to $6$~kHz (in 2022).
    \item \textbf{Debug stream}: events affected by online system failures, recovered automatically offline through a dedicated web-based application, currently under development.
\end{itemize}
With the only exception of the debug stream, this streaming model is \emph{inclusive}, such that a single event can be included in multiple streams should it satisfy the corresponding selection criteria.

\section{Summary and Outlook}

The ATLAS Trigger system successfully operates under the challenging conditions of Run~3, maintaining high efficiency and flexibility across a wide range of physics signatures. The upgrades to both the hardware (L1) and software (HLT) levels ensure robust performance in the presence of high pile-up, enabling a rich physics programme. Future developments will focus on preparing the Trigger system for the High-Luminosity LHC (HL-LHC) era, with enhanced hardware architectures, faster reconstruction algorithms, and advanced machine-learning techniques for real-time decision-making.


\end{document}